\documentclass[12pt]{iopart}
\usepackage{amsfonts}
\usepackage{iopams}
\usepackage{epsfig}
\usepackage{graphicx}
\usepackage{dcolumn}
\usepackage{bm}
\usepackage[T1]{fontenc}
\usepackage{ae,aecompl}

\begin{document}

\title{Shock Wave Response of Porous Materials: From Plasticity to Elasticity}
\author{Aiguo Xu, Guangcai Zhang, Yangjun Ying, Ping Zhang and Jianshi Zhu}
\address
{National Key Laboratory of Computational Physics, \\
Institute of Applied Physics and Computational Mathematics, P. O.
Box 8009-26, Beijing 100088, P.R.China\\
E-mail: Xu\_Aiguo@iapcm.ac.cn}
\date{\today}
\begin{abstract}
Shock wave reaction results in various characteristic regimes in
porous material. The geometrical and topological properties of these
 regimes are highly concerned in practical applications. Via the
morphological analysis to characteristic regimes with high
temperature, we investigate the thermodynamics of shocked porous
materials whose mechanical properties cover a wide range from
hyperplasticity to elasticity. It is found that, under fixed shock
strength, the total fractional area $A$ of the high-temperature
regimes with $T \geq T_{th}$ and its saturation value first
increase, then decrease with the increasing of the initial yield
$\sigma_{Y0}$, where $T_{th}$ is a given threshold value of
temperature $T$. In the shock-loading procedure, the fractional area
$A(t)$ may show the same behavior if $T_{th}$
 and $\sigma_{Y0}$ are chosen appropriately. Under the same
$A(t)$ behavior, $T_{th}$ first increases then decreases with
$\sigma_{Y0}$. At the maximum point $\sigma_{Y0M}$, the shock wave
contributes the maximum plastic work. Around $\sigma_{Y0M}$, two
materials with different mechanical properties may share the same
$A(t)$ behavior even for the same $T_{th}$. The characteristic
regimes in the material with the larger $\sigma_{Y0}$ are more
dispersed.

\end{abstract}


\maketitle

\section{Introduction}
Porous materials are ubiquitous in nature and extensively used as
industrial materials. Examples are referred to wood, bricks, metals,
foams, ceramics, carbon and explosives. The use of porous
 materials in parts may lead to reduced weight, improved structural
and mechanical properties, better heat transfer, greater motion and
deformation control, etc\cite{heterogeneous1,heterogeneous2}.
Besides others, they have also been used in surgical implant design
to fabricate devices to replace or augment soft and hard
tissues\cite{v2_1,v2_2}. Although the study on shock wave reaction
on porous materials has a long history. Most of previous studies
were focussed on the global behaviors, such as the
Hugoniots\cite{e1,e2,e3,e4,e5,e6,t1,t4} and the equation of
state\cite{eos1,eos2,eos3}. The dynamical procedures involve much
richer physical behaviors but in fact are much less extensively
investigated.

The main challenges for studying the dynamical behaviors in shocked
porous material are twofold: the first is the numerical tool, the
second is the scheme to analyze the simulation data. From the
simulation side, an appropriate simulation tool must overcome two
constraints. The first is the scale limitation. Molecular dynamics
can discover some atomistic mechanisms of shock-induced void
collapse\cite{Porous7,Yang}, but the spatial and temporal scales it
may cover are far from being comparable with experiments. To
overcome the scale limitation, one  solution is to develop some
mesoscopic particle methods. The second constraint is the numerical
stability. Traditional simulation methods, both the Eulerian and
Lagrangian ones, when treating with the dynamics of structured
and/or porous materials, encountered severe difficulties. The reason
is that material under investigation is generally highly distorted
during the collapsing of cavities. The Eulerian description is not
convenient for tracking interfaces. Lagrangian formulation has to
rezone the meshes to restore proper shapes. The mapping of state
fields of mass density, velocities and stresses from the distorted
mesh to the newly generated one is not easy and introduces errors.
In this study we use a mixed method, material-point
method(MPM)\cite{CTP2008,JPD2008}, to study shock wave reaction on
porous materials. As a step to approach the shock wave dynamics in
porous materials, we have carefully studied the cavity collapse in
shocked materials\cite{JPD2008,JPCM2007}.

As for the second challenge, data analysis and information picking
up, a relatively straightforward way is to study the local averaged
values and the corresponding fluctuations of state
variables\cite{CTP2009}. In Ref.\cite{CTP2009}, the evolution of
local turbulence mixing and volume dissipation were also studied.
Shock wave reaction results in various characteristic regimes in
porous materials, for example, regimes with high temperature,
regimes with high pressure, regimes with high particle speeds, etc.
These characteristic regimes are generally highly concerned in
practical applications. Regimes with high temperature are places
where initiation may start in energetic materials. Regimes with high
pressure are places where phase transition may occur. Regimes with
high particle speed are places where jet phenomena may occur. To
 understand the characteristic regimes defined
by $\Theta \geq \Theta_{th}$, we\cite{JPD2009} introduced the
Minkowski functionals to measure their morphological behaviors,
where $\Theta$ is a physical variable under consideration,  like the
temperature, density, some specific stress, particle velocity or its
components, $\Theta_{th}$ a given threshold value.

Previous studies showed that the porous metal
aluminum(Al)\cite{JPD2009} and porous HMX-like
material\cite{CTPHMX2009} show significant differences under shock
wave reaction. To clarify the effects of single material parameters
and present indicative information for material designs,
 it is interesting to have a
through study on the shock behaviors in relation to their mechanical
property ranging from hyper-plastic to pure elastic\cite{PRL2007}.

In present paper we focus on characteristic regimes with high
temperature. We organize the following part of the paper as follows.
Section 2 briefly introduces the material model and the numerical
method. Section 3 outlines the morphological characterization for
characteristic regimes manifested by Turing patterns. Simulation
results are shown and analyzed in section 4. Section 5 makes the
conclusion.

\section{Material model and material-point method}

The porous material is fabricated by a solid body with a number of
randomly distributed voids embedded. The solid body follows an
associative von Mises plasticity model with linear kinematic and
isotropic hardening\cite{CTP2009}.
 The pressure $P$ is
calculated by using the Mie-Gr\"{u}neissen equation of state which
can be written as $ P-P_{H}=[\gamma (V)/V][E-E_{H}(V_{H})]$, where
$P_{H}$, $V_{H}$ and $E_{H}$ are pressure, specific volume and
energy on the
Rankine-Hugoniot curve, respectively. The relation between $P_{H}$ and $%
V_{H} $ can be estimated by experiments and be written as
\begin{equation}
P_{H}=\left\{
\begin{array}{ll}
\frac{\rho _{0}c_{0}^{2}(1-\frac{V_{H}}{V_{0}})}{(\lambda -1)^{2}(\frac{%
\lambda }{\lambda -1}\times \frac{V_{H}}{V_{0}}-1)^{2}}, & V_{H}\leq
V_{0}
\\
\rho _{0}c_{0}^{2}(\frac{V_{H}}{V_{0}}-1), & V_{H}>V_{0}%
\end{array}%
\right.
\end{equation}
where $\rho_0$, $V_0$ are the initial density and specific volume of
the solid material, $c_0$ the sound speed, $\lambda$ the coefficient
in Hugoniot velocity relation. Both the shock compression and the
plastic work $E-E_{H}(V_{H})$ cause the increasing of temperature.
The increasing of temperature from shock compression can be
calculated as:
\begin{equation}
\frac{\mathrm{d}T_{H}}{\mathrm{d}V_{H}}=\frac{c_{0}^{2}\cdot \lambda
(V_{0}-V_{H})^{2}}{c_{v}\big[(\lambda -1)V_{0}-\lambda V_{H}\big]^{3}}-\frac{%
\gamma (V)}{V_{H}}T_{H}.  \label{eq-eos-temprshock}
\end{equation}%
where $c_{v}$ is the specific heat. Eq.(\ref{eq-eos-temprshock}) can
be derived from the thermal equation and the Mie-Gr\"{u}neissen
equation of state\cite{CTP2009}. The increasing of temperature from
plastic work can be calculated as: $\mathrm{d}T_{p}=\mathrm{d}W_{p}/
c_{v}$.

We model shocked materials with continuously varying mechanical
properties. The reference material is the metal aluminum. The
corresponding parameters are as below: density in solid portion
$\rho_{0}=2700$ kg/m$^{3}$, Yang's module $E=69$ Mpa, Poisson's
ratio $\nu =0.33$, initial yield $\sigma _{Y0}=120$ Mpa, tangential
module $E_{\tan }=384$ MPa, sound speed $c_{0}=5.35$ km/s,
characteristic coefficient in the Hugoniot velocity relation
$\lambda=1.34$, specific heat $c_{v}=880$ J/(Kg$\cdot $K), heat
conduction coefficient $k=237$ W/(m$\cdot $K), and Gr\"{u}neissen
coefficient $\gamma=1.96$. The initial temperature of the material
is 300 K. In simulations a wide range of the yield $\sigma _{Y0}$
will be used.

The material point method is a relatively new particle method in
computational solid mechanics. This method uses a regular structured
grid as a computational scratchpad for computing spatial gradients
of field variables. The grid is convected with the particles during
deformations that occur over a time step, eliminating the diffusion
problems associated with advection on an Eulerian grid. The grid is
restored to its original location at the end of a time step. In
addition to avoiding the Eulerian diffusion problem, this approach
also circumvents problems with mesh entanglement that can plague
fully Lagrangian-based techniques when large deformations are
encountered. The MPM has also been successful in solving problems
involving impact, etc. It has an advantage over traditional finite
element methods in that the use of the regular grid eliminates the
need for doing costly searches for contact surfaces. Details of the
scheme is referred to our previous
publications\cite{JPD2008,CTP2008}.

\section{Outline of morphological description}

A variety of techniques can be used to describe the complex spatial
distribution and time evolution of physical quantities in the
shocked porous material. In this study we concentrate on the set of
statistics known as Minkowski
functionals\cite{Minkowski1903,new1,JSTAT2008}. The Minkowski
functionals has been successfully used to characterize patterns in
reaction-diffusion systems\cite{PRE1996}, spinodal
decomposition\cite{PRE1997,JCP2000}, fluctuations of cosmic
microwave background\cite{PRD2004}, block copolymer
systems\cite{JCP2004,PRE2008} and to reconstruct complex
materials\cite{PRL2003}.

Assume $\Theta$ is a physical quantity being interesting to us, then
the regions with $\Theta \geq \Theta_{th}$ in the shocked porous
material are referred to as characteristic regimes, where
$\Theta_{th}$ is a threshold value of $\Theta$. To simplify the
analysis of the complex physical field, we first condense the
physical field as two kinds of characteristic regimes, the white and
 the black. The white correspond to regimes with $\Theta \geq
\Theta_{th}$ and the black correspond to regimes with $\Theta <
\Theta_{th}$. For such Turing patterns, a general theorem of
integral geometry states that all properties of a $d$-dimensional
convex set which satisfy motion invariance and additivity (called
morphological properties) are contained in $d+1$ numerical values
\cite{Hadwiger19561959}. For a condensed temperature field, the
white correspond to the high-temperature regimes and the black
correspond to the low-temperature regimes. The high-temperature
regimes are also generally referred to ``hot-spots".

For a two-dimensional temperature map, the three Minkowski
functionals correspond geometrically to the total fractional area
$A$ of the high-temperature regimes, the boundary length $L$ between
the high- and low-temperature regimes regimes per unit area, and the
Euler characteristic $\chi $ per unit area (equivalent to the
topological genus). When we increase the temperature threshold
$T_{th}$ from the lowest temperature to the highest one, the
high-temperature area $A$ will decrease from $1$ to $0$; the
boundary length $L$ first increases from $0$, then arrives at a
maximum value, finally decreases to $0$ again. There are several
ways to define the Euler characteristic $\chi $. Two simplest one is
$\chi = (N_{W}-N_{B})/N$, where $N_{W}$ ($N_{B}$) is the number of
connected white (black) regimes, $N$ is the total number of pixels.
In contrast to the area $A$ and boundary length $L$, the Euler
characteristic $\chi $ describes the connectivity of the
characteristic regimes in the material. It describes the patterns in
a purely topological way, i.e., without referring to any kind of
metric. It is negative (positive) if many disconnected black (white)
regimes dominate the image. A vanishing Euler characteristic
indicates a highly connected structure with equal amount of black
and white regimes. The ratio $ \kappa = \chi/L$ describes the mean
curvature of the boundary line separating black and white regimes.
More discussions and calculation schemes of the Minkowski
functionals are referred to Refs.\cite{new1,PRE1996,new3}.

  Among the three Minkowski functionals, the high-temperature area $A$
is the only one which monotonically increases in the shock-loading
procedure and/or when the threshold value becomes smaller. For a
given $T_{th}$, its increasing rate, $D$, presents meaningful
information. When the temperature threshold $T_{th}$ becomes higher,
$D$ decreases. The variations of $A$, $L$ and $\chi$ with $T_{th}$
and time $t$ compose a scenario for the shock response of porous
material\cite{JPD2009}.

\begin{figure}[tbp]
\caption{(Color online) Snapshots of shocked porous material, where
$\Delta = 0.5$, $v_{init}=1000$ m/s, t=1000 ns, . From left to right
and from top to bottom, the initial yields in the two rows are
$\sigma_{Y0} =$12, 120, 3000, 8000, 10000, 12000, 15000, 20000 Mpa,
respectively. The unit of length is 10$\mu$m. The width and height
of the simulated system are 1mm and 5mm, respectively. From blue to
red the color corresponds to increase of temperature.}
\end{figure}


\section{Simulation results and physical interpretation}

If denote the mean density of the porous body as $\rho $, then the
porosity of the material is $\Delta =1 -
\rho/\rho_{0}$\cite{definition}. The shock  to target body is loaded
 by colliding with a rigid wall being static at the bottom position
$y=0$.  The initial velocity of the porous body is $-v_{init}$. The
collision starts at time $t=0$. The height and width of the porous
body are 5mm and 1mm, respectively. Periodic boundary conditions are
used in the horizontal directions, which means that the real system
is composed of many of the simulated ones aligned periodically in
the horizontal direction.

\begin{figure}[tbp]
\centerline{\epsfig{file= 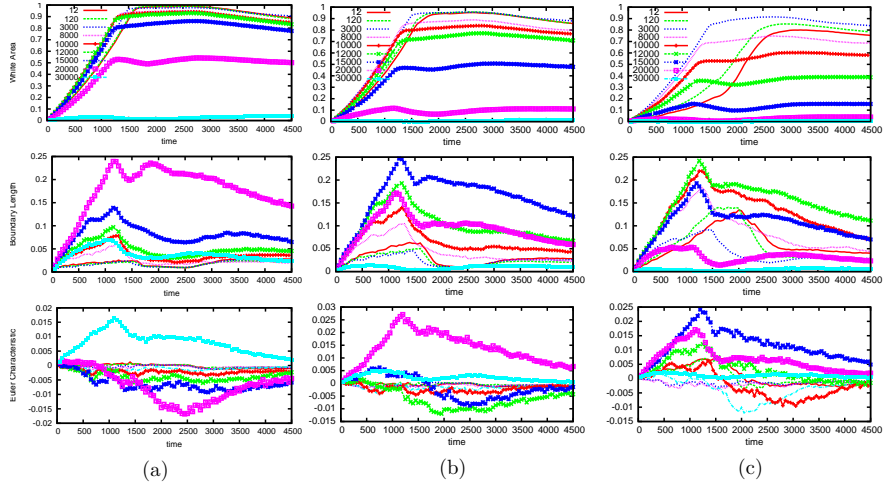, 
bbllx=104 pt,bblly=330 pt,bburx=533 pt,bbury=571 pt,
width=0.75\textwidth,clip=}} \caption{(Color online) Minkowski
measurements for cases with various initial yields, where  $\Delta =
0.5$, $v_{init}=1000$ m/s. The values of $\sigma_{Y0}$ are shown in
the legend with the unit MPa. $T_{th}=$ 400K in (a), $T_{th}=$ 500K
in (b), $T_{th}=$ 600K in (c). }
\end{figure}

\subsection{Case with $v_{init} = 1000$ m/s}

Figure 1 presents a series of snapshots in the shock loading
procedure, where the porosity $\Delta=0.5$ and initial velocity
$v_{init}=1000$m/s, the time $t=$1000ns. From left to right and from
top to bottom, the two rows of snapshots are for cases with
$\sigma_{Y0}=$12, 120, 3000, 8000, 10000, 12000, 15000, 20000 Mpa,
respectively.
 The points in the figure correspond to the
 material particles. The color from blue to red corresponds to
 increasing of temperature. From Fig.1 one can find the moving down
 of upper free surface and moving up of global compressive-waves series.
  For each case, the initial
shock wave is decomposed as a complex series of compressive and
rarefactive waves. In the shock loading procedure, the compressive
effects dominate. Within the shocked region, both the plastic work
and shock compression make the temperature increase. In the present
case, the plastic work dominates. The high-temperature regimes in
materials with higher initial yields are more dispersed. The
increasing rate of high-temperature area $A(t)$ is different when
the initial yield changes.

 To quantify
and get a more complete understanding on the shock wave response
behavior, we show a set of morphological measures versus time in
Fig.2(a), where values of initial yields are shown in the legend.
 The temperature threshold here is $T_{th}= $400 K.  With the
propagation of compressive-wave-series in the porous material, the
 high-temperature area $A$ first increases
with time in a parabolic way, then approaches to a saturation value
slowly, keeps the saturation value for a period, finally decreases.
The final decrease indicates that rarefactive waves are reflected
back from the upper free surface and are decreasing the mean
pressure and temperature, an
 amount of material particles changed their temperature from $T >
 T_{th}$ to $T < T_{th}$. During the saturation period, more
 compressive waves arrived at the upper free surface, consequently,
 more rarefactive waves are reflected back into the porous body.
 The former tends to increase the temperature, while the latter tends
  to decrease. The two effects are nearly balanced. Therefore,
  the high-temperature area $A$ keeps nearly a constant.

As for effects of the initial yield $\sigma_{Y0}$, we can find two
interesting phenomena: Both the initial increasing rate and the
saturation value of $A$ first increase, then decrease when the
material changes from being superplastic to pure elastic. For cases
checked in our numerical experiments, the increasing rate $D$
becomes larger when the initial yield $\sigma_{Y0}$ increases from
$0$ to about $10$ Gpa. If $\sigma_{Y0}$ further to increase, $D$
will decrease. The saturation value $A_S$ of high-temperature area
becomes larger when $\sigma_{Y0}$ increases from being very small to
about $1$ Gpa. When $\sigma_{Y0}$ becomes larger, the saturation
value $A_S$ decreases. For the time interval shown Fig. 2(a), when
$\sigma_{Y0} > 10$ Gpa, the saturation value $A_S$ decreases and the
period for $A \approx A_S$ becomes shorter with the increasing of
$\sigma_{Y0}$. When $\sigma_{Y0} = 12$ Gpa, $A \approx 0.96 $ during
the period $1430$ ns $< t < 2924$ ns; When $\sigma_{Y0} = 15$ Gpa,
$A$ increases slowly from $0.8$ to $0.86$ during the period $1245$
ns $< t < 2825$ns; When $\sigma_{Y0} = 20$ Gpa, $A_S \approx 0.52$
and have a local minimum value $0.49$ at about $t = 1863$ns. When
$\sigma_{Y0} = 30$ Gpa, the area $A$ keeps very small.

Now we check the information given by the boundary length and Euler
characteristic in Fig.2(a). For all cases shown in the figure, the
boundary length $L$ first increases, then decreases with time. The
former increase corresponds to the propagation of compressive waves
and the appearance of more hot-spots. The decreasing of $L(t)$ is
not monotonic. The initial decrease corresponds to the coalescence
of some hot-spots, the latter decrease corresponds to the coming in
of the global rarefactive waves from the upper free surface, which
result in expanding and coalescence of some cold-spots. A prominent
behavior here is that, for the case with $\sigma_{Y0} = 20$ Gpa,
 the boundary length $L$  has the largest value. For this case, the value of $\chi$
changes from being slightly positive to being the most negative.
Combining information of $A$, $L$ and $\chi$, we can know that, with
the propagation of compressive wave in the porous material, the
number of hot-spots with $T > 400 $K quickly increases, but
distributes quite scatterredly. After corresponding compressive wave
scanned all the material body, some scattered cold-spots with $T <
400 $K expand and partly coalescence due to the coming in of
rarefactive waves. During this procedure, some small hot-spots with
$T > 400 $K disappear. Therefore, both the high-temperature area $A$
and boundary length $L$ decrease and the Euler Characteristic $\chi$
becomes more negative. For other cases, the smaller the boundary
length $L$, the flatter the wave front in the temperature map.
 The material with $\sigma_{Y0} = 15$ Gpa has the secondary maximum
boundary length $L$ and more flatter $\chi(t)$ curve than the
material with $\sigma_{Y0} =20$ Gpa. This means that the numbers of
hot-spots and cold-spots do not have much difference. The number of
cold-spots dominates slightly during the time interval shown in the
figure.

In the shock loading procedure,  if we decrease the threshold value
$T_{th}$, the wave fronts in the pixelized temperature map becomes
flater. Consequently, the values of $L(t)$  are smaller,
 the $\chi (t)$ values are closer to zero, and the curves for $A(t)$
 becomes closer to be linear.
 If we increase the threshold value $T_{th}$, the pixelized temperature
 map shows different geometric and topological  behaviors. Examples are referred to Figs. 2(b) and 2(c).
 $T_{th} = 500 $K in Fig. 2(b) and $T_{th} = 600 $K in Fig. 2(c).
It is clear that  the saturation values of $A(t)$ decrease with the
increasing of $T_{th}$. In the shock-loading procedure, the material
with $\sigma_{Y0} = 12$ Mpa has about $20\%$ of material particles
can not get a temperature higher than 600K and $5\%$  can not get
the temperature higher than 500K, and only $1\%$ can not
 get the temperature higher than 400K. For
the material with $\sigma_{Y0} = 120$ Mpa, in the shock loading
procedure, there are about $15\%$ of material particles can not get
the temperature higher than 600K, $4\%$  can not get the temperature
higher than 500K, and only $1\%$  can not get the temperature higher
than 400K.

 When the initial yield is
very high, for example, $\sigma_{Y0} = 15$Gpa, the material is very
elastic. Consequently, the saturation value $A_{S}$ of
high-temperature area is small. For example, $A_S = 0.15$ when
$T_{th} = 600$K, which means that $85\%$ of material particles can
not get the temperature higher than 600K. For the case with
$\sigma_{Y0} = 20$Gpa, $A_S = 0.54$ when $T_{th} = 400$K, $A_S =
0.11$ when $T_{th} = 500$K and $A_S = 0.04$ when $T_{th} = 600$K.
For the case with $\sigma_{Y0} = 30$Gpa, only $0.1\%$ of material
particles can get a temperature higher than 600K in the shock
loading procedure. In the temperature pattern with $T_{th}=400$K,
the case of $\sigma_{Y0} = 20$GPa has a largest boundary length.
When $T_{th}=500$K, the case of $\sigma_{Y0} = 15$GPa has a largest
boundary length. When $T_{th}=600$K, the case of $\sigma_{Y0} =
10$GPa has the largest $L$.

\begin{figure}[tbp]
\centerline{\epsfig{file= 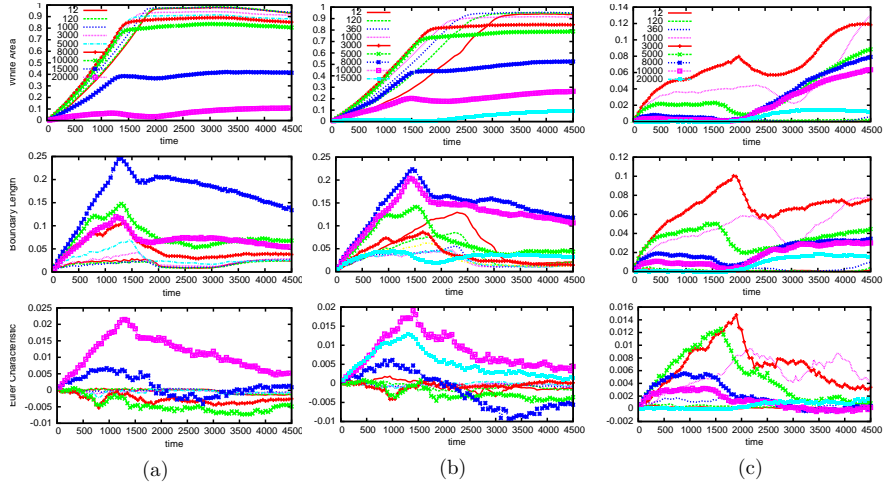, 
bbllx=104 pt,bblly=330 pt,bburx=533 pt,bbury=571 pt,
width=0.75\textwidth,clip=}} \caption{(Color online) Minkowski
measurements for cases with various initial yields, where  $\Delta =
0.5$. The values of $\sigma_{Y0}$ are shown in the legend with the
unit MPa. The threshold temperature is 400K. $v_{init}=800$ m/s in
(a), $v_{init}=600$ m/s in (b),$v_{init}=400$ m/s in (c).}
\end{figure}

\begin{figure}[tbp]
\centerline {\epsfig{file=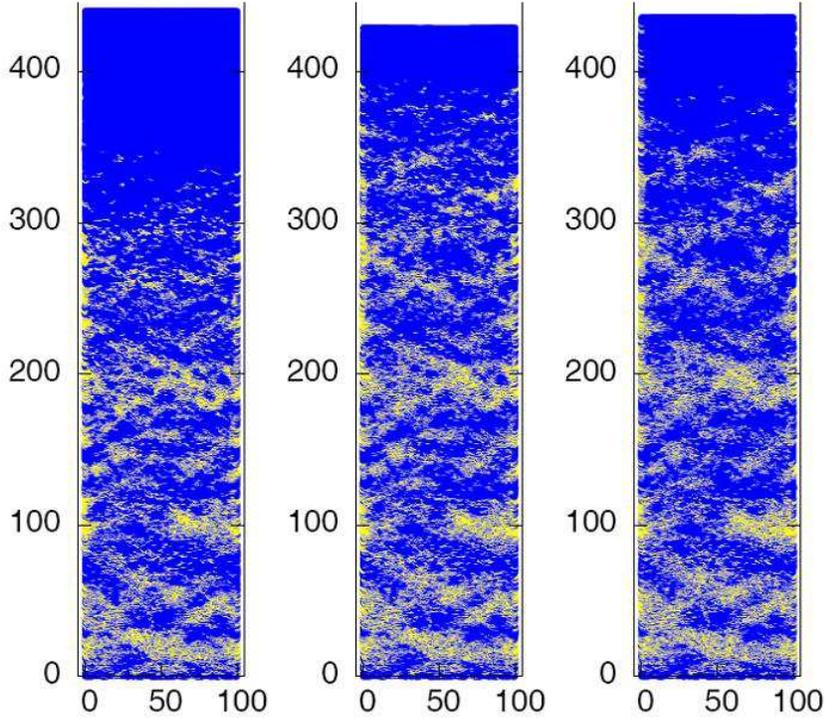,
bbllx=90pt,bblly=50pt,bburx=612pt,bbury=511pt,
width=0.7\textwidth,clip=}} \caption{(Color online) Configurations
with condensed temperature map, where $\Delta = 0.5$, $v_{init}=$400
m/s, $\sigma_{Y0}=$3000 Mpa. The areas with temperature higher than
400K are shown in yellow, other areas are in blue. The times
corresponding to the three snapshots  are 1500ns, 2000ns and 2500ns,
respectively, from which one can find that the high-temperature area
at t=2000ns is a little larger than that at t=2500ns and that at
t=1500ns. }
\end{figure}
\begin{figure}[tbp]
\centerline{\epsfig{file= 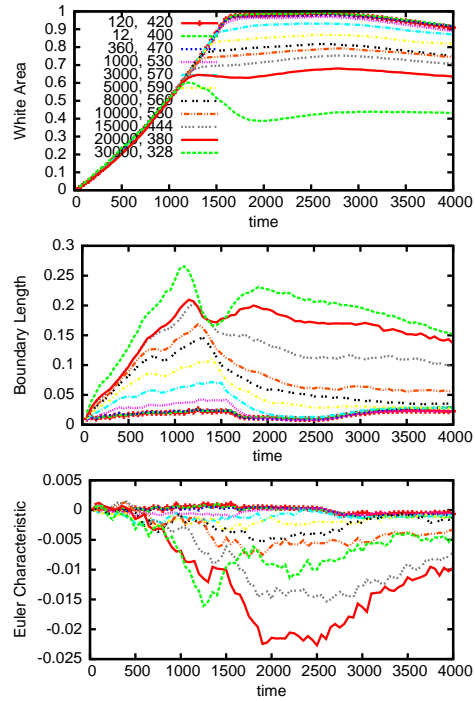, 
bbllx=104 pt,bblly=96 pt,bburx=513 pt,bbury=705 pt,
width=0.4\textwidth,clip=}} \caption{(Color online) Minkowski
measurements versus time, where $\Delta = 0.5$, $v_{init}=1000$ m/s,
the values of $\sigma_{Y0}$ and $T_{th}$ are shown in the two
columns of the legend. The units are MPa and K, respectively.}
\end{figure}
\begin{figure}[tbp]
\centerline {\epsfig{file= 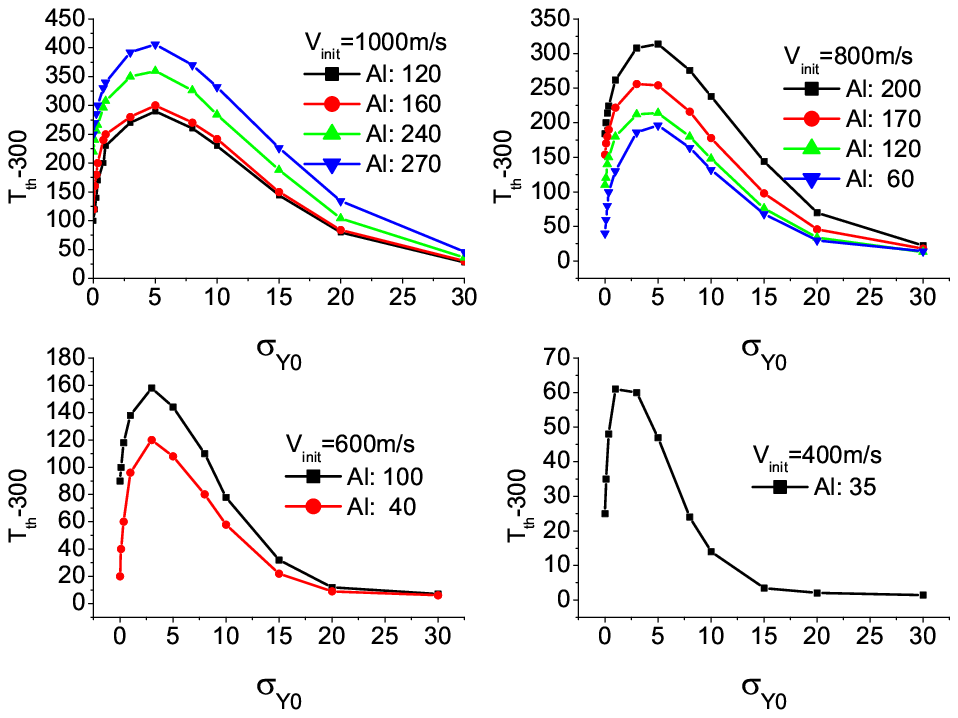, width=0.75\textwidth,clip=}}
\caption{(Color online) $T_{th}$ versus $\sigma_{Y0}$ for the same
$A(t)$ behavior. The initial impact velocities are shown in the
legends. Each curve in the figure is labeled by the $T_{th}$ value
for the reference material, aluminum (Al). The units for the
horizontal and vertical axes are GPa and K, respectively. }
\end{figure}

\subsection{Effects of initial shock strength}

With the decreasing of initial shock strength, both the highest and
the mean temperatures in the shocked portion decrease. A set of
Minkowski measures for the shocking procedure of porous materials
with various initial yields are shown in Fig.3, where $\Delta = 0.5$
and $T_{th}$=400K.  The initial impact velocities are different in
Figs. (a), (b) and (c).  They are 800m/s, 600m/s and 400m/s,
respectively. The values of initial yields are shown in the legends.
Specifically,  the initial yields are 12, 120, 1000,3000, 5000,
8000, 10000, 15000 and 20000MPa in Fig.3(a) ; in Fig.3(b) they are
12, 120, 360, 1000,3000, 5000, 8000, 10000 and 15000MPa; in Fig.3(c)
they are 12, 120, 360, 1000,3000, 5000, 8000, 10000, and 20000MPa.
Compared with cases shown in Fig. 2(a), both the saturation value
$A_S$ and the increasing rate $D(T_{th})$ of the high-temperature
area decrease when the initial shock becomes weaker. As an example,
for the material with $\sigma_{Y0}=$1000Mpa, when the initial impact
velocity is $v_{init}=$1000m/s, there are 99\% of material particles
arrive at the temperature higher than 400K in the shock-loading
procedure; when $v_{init}=$800m/s, the fraction of material
particles getting the temperature higher than 400K is 97\%; when
$v_{init}=$600m/s, the fraction is 89\% ; when $v_{init}=$400m/s,
the fraction becomes only 5\%.

In Fig.3(c), the case of $\sigma_{Y0}=$3000 MPa has the maximum
high-temperature area $A$, boundary length $L$ and Euler
characteristic $\chi$. In this figure, only after the
high-temperature area $A$ gets its maximum value, it begins to
decrease, which means that, under such a shock strength,
 most material particles  can not get a temperature higher than 400K,
 the threshold
value 400K has been very close to the highest temperature in this
system. To understand better why the maximum high-temperature area
occurs at about t=2000ns, we show the configurations with condensed
temperature map in Fig.4, where the three snapshots are for the
times t=1500ns, 2000ns and 2500ns, respectively. The areas with
temperature higher than 400K are shown in yellow, other areas are
shown in blue. One can find that the high-temperature area at time
t=2000ns is the largest among the three snapshots. From the heights
of the upper free surface, one can find that the one for t=2000ns is
the lowest, which means that the shock-loading procedure finishes at
about that time. When the unloading procedure starts, the area with
high-temperature decreases. From Fig.4 one can also find that the
high-temperature regimes for $T_{th}=400$K have been very dispersed.
This is consistent with the large value of boundary length, and
consistent with the above observation that only a small portion of
material particles get the temperature higher than 400K under such a
shock strength.

\subsection{($\sigma_{Y0}$, $T_{th}$) pairs for the same $A(t)$ behavior}

As mentioned above, among the three Minkowski functionals, the
high-temperature area $A(t)$ is the only one being monotonic when
the threshold value $T_{th}$ decreases and/or with the going on of
the shock-loading procedure. It is natural to check if $A(t)$ shows
the same behavior when using appropriate ($\sigma_{Y0}$, $T_{th}$)
pairs. Figure 5 shows such examples for the case with initial impact
velocity $v_{init}=$1000m/s. In Fig.5, the temperature threshold for
reference material, aluminum,  is 420K. From Fig. 5 we can find that
 materials with different initial yields, if we choose an
appropriate $T_{th}$ to observe, the high-temperature area $A(t)$
shows the same behavior in the shock-loading procedure. Such a
property can be understood better by observing Figs. 1 and 4. From
Figs. 1 and 4, it is also clear that, for a fixed shock strength,
with the increasing of the initial yield, the wave front becomes
wider, the high-temperature regimes becomes more scattered, more
low-temperature domains are embedded in the compressed portion. This
morphological
 behavior is manifested by larger boundary lengths  and more negative
 Euler characteristics in Fig.5.

If use the ($\sigma_{Y0}$, $T_{th}$) pairs in Fig.5 as coordinates,
we get the curve labeled by ``$V_{init}$=1000m/s" and ``Al: 120"  in
Fig.6. For the shock strength $v_{init}=$1000m/s, the material with
$\sigma_{Y0}=$5GPa has the maximum $T_{th}$ which is about 590K. In
this case, the shock contributes the maximum plastic work and the
system has the highest temperature. If increase the temperature
threshold $T_{th}$ to 460K, 540K and 570K, we have the other three
curves.
 Along each of them, $A(t)$ shows the same behavior in the
shock-loading procedure. If decrease the shock strength to
$v_{init}=$800m/s, 600m/s and 400m/s, we get the other three plots
of Fig.6. The maximum value point $\sigma_{Y0M}$ moves towards the
lower value of the initial yield when the shock wave becomes weaker.
In the shock loading procedure,  if we decrease the threshold value
$T_{th}$, the wave fronts  becomes flater. Consequently,  $L(t)$
curves become smaller, $\chi (t)$ becomes closer to zero; the curves
for $A(t)$ becomes closer to be linear.

\section{Conclusion}

Shock wave reaction results in various characteristic regimes in
porous materials. The properties of these regimes are highly
concerned in practical applications. Based on the material-point
simulation and morphological characterization, we investigate how
the initial yield influence the behavior of high temperature regimes
in shocked porous material.
 It is found
that, under fixed shock strength, the total fractional area $A$ of
high-temperature regimes (with $T \geq T_{th}$) and its saturation
value first increase, then decrease when initial yield $\sigma_{Y0}$
becomes higher. In the shock-loading procedure the fractional area
$A(t)$ may show the same behavior under various choices of $T_{th}$
and $\sigma_{Y0}$.
 For the same $A(t)$ behavior, $T_{th}$ first increases then decreases when
$\sigma_{Y0}$ becomes higher. At the maximum point $\sigma_{Y0M}$,
the plastic work by the shock gets the maximum value. Around
$\sigma_{Y0M}$, two materials with different mechanical properties
may share the same $A(t)$ behavior even for the same threshold
$T_{th}$. The high-temperature regimes in the material with the
higher initial yield $\sigma_{Y0}$ are more dispersed. Other kinds
of characteristic regimes, for example, those with high pressure,
high particle speed, etc., can be studied in the same way.

\section*{Acknowledgments}
A. Xu is grateful to Prof. Hua Li for helpful discussions on shock
waves and porous materials, to Drs. G. Gonnella, A. Lamura and V.
Sofonea for helpful discussions on morphological description. This
work is supported partly by Science Foundations of Laboratory of
Computational Physics, China Academy of Engineering Physics [under
Grant Nos. 2009A0102005, 2009B0101012], and National Science
Foundation of China (under Grant Nos.10702010, 10775018
 and 10604010).


\begin{thebibliography}{99}
\bibitem{definition} In our previous paper\cite{JPD2009}, the
porosity is described by the ratio $\delta = \rho_0/\rho$. The
relation between $\Delta$ and $\delta$ is $\Delta = 1 - 1/\delta$.
The larger $\delta$ is, the larger $\Delta$ is.
\end{thebibliography}
\end{document}